\def\year{2017}\relax
\documentclass[letterpaper]{article}
\usepackage{aaai17}
\usepackage{times}
\usepackage{helvet}
\usepackage{courier}
\usepackage{multicol}
\usepackage{graphicx}
\usepackage{amsmath}
\usepackage{epstopdf}
\begin{document}
%
\title{Identifying Unsafe Videos on Online Public Media using Real-time Crowdsourcing}
\author{Sankar Kumar Mridha$^1$, Braznev Sarkar$^1$, Sujoy Chatterjee$^2$ and Malay Bhattacharyya$^1$\\
$^1$Department of Information Technology, Indian Institute of Engineering Science and Technology, Shibpur\\ Howrah -- 711103, India\\
E-mail: \{msankar, braznev.rs2015, malaybhattacharyya\}@it.iiests.ac.in\\
$^2$Department of Computer Science and Engineering, University of Kalyani, Nadia -- 741235, India\\
E-mail: sujoy@klyuniv.ac.in\\
}
\maketitle

\begin{abstract}
Due to the significant growth of social networking and human activities through the web in recent years, attention to analyzing big data using real-time crowdsourcing has increased. This data may appear in the form of streaming images, audio or videos. In this paper, we address the problem of deciding the appropriateness of streaming videos in public media with the help of crowdsourcing in real-time.
\end{abstract}

\section{Introduction}
Inspection of streaming data in real-time is a challenging task. Streaming data (be it in the form of images, audio or videos) are required to be analyzed on the fly because the data can not be stored for a longer period \cite{Muthukrishnan2005}. In reality, for the sake of simplicity, streaming videos are occasionally considered in a semi-streaming setting \cite{Galasso2014}. We can assume that streaming videos are captured as a set of segments for a better processing. Day by day, the global activity through the Web is increasing in daily life. The task of monitoring unlawful activities is becoming more and more crucial. Detecting the illegitimate portions (unsafe) in the streaming videos with the help of artificial intelligence is time consuming and may not be suitable in many circumstances. Alternatively, crowdsourcing can be used efficiently and effectively for such purposes \cite{Yeung1998}.

Crowdsourcing is a model of virtual platforms where the crowd can share their views, innovative ideas, skills, etc. In crowdsourcing, any kind of micro-task can be solved in real-time by the crowd workers creating an online labor market \cite{Ho2012,Gonen2014}. Processing of crowd data might also require other domain expertise like social network analysis \cite{Doan2011}, judgment analysis \cite{Chatterjee2017}, etc. The workers either volunteer or get remuneration for solving the tasks. So, we can make judgments on images, audio or videos (even in streaming form) available in public media by employing the crowd workers in real-time without or through remuneration. The people inclined to the online media like YouTube, Facebook, Twitter, etc. can voluntarily provide support in this regard. In this paper, we propose an approach that can detect unsafe (as per the general policy of online contents) videos available online by the help of active crowd workers in a minimal time. To judge the quality of workers, we can evaluate their performance based on submitted solutions. When a set of workers submit their opinions for a video segment, final judgment can be made by the aggregation of their opinions. Based on this, we can make the final decision about the video. We also discuss the evident challenges like diversity of the video contents, which suggest to include the demographic information of crowd workers in the model \cite{Liu2012}.

\section{Motivation}
There is a significant growth in social networking and human activities online in recent years. As a result, an enormous amount of data is getting included in the online public media. Many real-life applications demand for analyzing such contents in real-time before making them public. We are focusing on streaming videos to judge whether they satisfy the policies of respected applications or not. A large number of people are always engaged in different activities (e.g., video search, watching videos, etc.) online. We propose to involve these persons as crowd workers and take their opinions in deciding the appropriateness of video contents. It is considerable that a worker can take part in a short job at the time of performing his won work. So, we can post a segment of video, with a short time-frame, as a short job. Considering the YouTube platform, we demonstrate the working principle of our proposed model. As per the latest statistics, YouTube experiences a video upload rate of 5hr/second (on an average). Furthermore, the average number of videos that are watched in each second on YouTube is more than 57k. The (minimum) number of visitors who watch these videos in YouTube is about 350 per second. Mobile users also spend a long time (40min on average) watching videos on YouTube. These facts highlight a huge potential of using the viewers as feedback providers for verifying the appropriateness of uploaded videos in real-time.

\section{Proposed Method}
\begin{figure}[!t]
\begin{center}
\includegraphics[width=3cm]{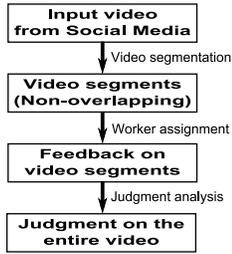}
\caption{Flowchart of the proposed model.}
\label{Fig:Flowchart}
\end{center}
\end{figure}

Let us consider that a streaming video (or a complete video) $V$ is uploaded by a user. The video $V$ is divided into $n$ segments (containing a set of frames) represented as $\{v_1, v_2, \ldots, v_n\}$ ($\bigcup v_i = V$ and $v_i \cap v_j = \phi$ for all $i$, $j$). Each video segment has a minimum running time of $\tau$ (to judge whether it is unsafe). Each segment is judged by a set of workers represented as $W_i = \{w_{i1}, w_{i2}, \ldots, w_{i{m_i}}\}$, for all $i$, and $m_i$ denotes the number of workers for the $i^{th}$ segment. Note that, the set of workers for each segment are not necessarily distinct. The workers for each segment provide their opinions in the form of Yes (Y) or No (N), which signify whether the video segment is accepted or rejected as being safe, respectively. After receiving the responses on each segment, decision is taken by majority voting. So, for each segment we get a judgment either as Yes or No. The video will be considered safe (suitable) if all of segments are judged as Yes, otherwise not. A schematic view of the entire approach is shown in Fig.~\ref{Fig:Flowchart} and is detailed hereunder.

\textbf{Segmentation of videos:} The foremost step is to decompose a video into minimum time slots so that it contains sufficient amount of information for making a judgment. We adopt a recent approach of Deza et al. explaining the way to improve the human performance in a realistic sequential visual search task \cite{Deza2017}. We consider the (minimum) time duration $\tau$ required for understanding the video segment by the workers.

\textbf{Assigning crowd workers:} Thereafter comes the proper selection of workers for assigning the video segments. If a segment is judged by a single worker then the outcome is not always reliable, so we assign a set of workers to solve a common segment independently. Online viewers are of two types -- signed (mobile users and others who enter into the platform with unique identity) and unsigned (whose profile is unknown). Primarily, we prefer the signed viewers as the crow workers to judge the segments. If signed workers are not sufficient, then we will consider the unsigned workers. Based on the performance, we are awarding the workers with credit points to motivate them. We also track the workers' profile for a better assignment. If a video is posted with American accent then it is obvious that it will be perfectly perceived by a worker who is native American.

\textbf{Judgment analysis on each segment:} As the videos are coming in a streamline, the accuracy of crowd workers are computed with respect to a fixed sized sliding window. Majority voting is initially applied to a particular video and based on the agreement of worker's opinion with the majority, the accuracy of a particular crowd worker is calculated. These accuracy values are then incorporated in the final stage while aggregating the opinions. As it is obvious that accuracy of a worker is not constant for different time instants, therefore the accuracy scores are tracked after certain time intervals. Thus, majority voting is repeatedly applied for a particular video within particular time interval. The other issue that we consider is the biasness of workers over a particular video. Biasness of crowd workers means while providing their opinions they might be inclined to either Yes or No. To identify these characteristics, some videos (as test cases) having ground truth label are included and based on the responses on them the biasness is determined.

\textbf{Final decision making:} Final decision is taken based on the collective judgment on all the segments. If all the segments are marked Yes then it means that the entire video is acceptable by the workers. So, the video is safe for the platform. If any of the segment is rejected by the assigned group, then the video is unsafe.

\section{Challenges}
Every second massive amount videos get uploaded by the different users. Judging a streaming video in real-time by the crowd workers is a demanding job. It depends on many factors such as the selection of workers, fixing the number of workers for each segment \cite{carvalho2016}, response time of each worker, duration of each video segment, etc. It is also challenging to find the ways to segregate a video into overlapping segments for reducing the misinterpretation at the time of judgement. Attributes of a video and demography of the assigned workers play a major role in evaluating the judgment too. So, the same video may not be equally suitable for a pair of workers with separate demography. To better understand the other challenges, we took a survey on 45 daily viewers of YouTube of whom 71\% were willing to be crowd workers. However, most of them (73\%) were ready to provide only a binary response, not the detailed feedback. On an average, they preferred a video segment size of about 140sec and a gap of 37min between the arrival of two successive segments.

\section{Conclusion}
The proposed model is a good fit for the YouTube videos because they can be checked as and when uploaded by the users to determine whether they satisfy the safety policies. This model can also be implemented directly in case of streaming videos received from CCTVs for detecting abnormal activities in surveillant area. The approach is also applicable towards detecting the suitability of images and audio, when they get uploaded to other online social media satisfying their policies. Unlike the existing approaches, where videos are blocked with a delay after the users complain, the proposed one urges a real-time action. However, there remain some ethical challenges like managing the risks involved in having contact with potentially harmful media and their impact on the relationship with the workers.

\section{Acknowledgment}
The work of Malay Bhattacharyya is supported by the Visvesvaraya Young Faculty Research Fellowship 2015-16 of MeitY, Government of India. All the authors would like to thank the crowd contributors involved in this work.

\bibliographystyle{aaai}
\bibliography{Reference}

\end{document}